\documentclass{article}
\usepackage[dvips]{graphicx}
\usepackage{amssymb}
\usepackage{amsmath}
\usepackage{amscd}
\usepackage{mathrsfs}
\usepackage{latexsym}

\date{}
\author{Claudio Garola\footnote{E-mail: Garola@le.infn.it} \ \ and \ Sandro Sozzo\footnote{E-mail: Sozzo@le.infn.it} 
\\ Dipartimento di Fisica and Sezione INFN \\ Universit\`a del Salento, via Arnesano, 73100 Lecce, Italy}

\title{\textbf{Recovering Quantum Logic within an Extended Classical Framework}}

\begin{document}

\maketitle

\begin{abstract}
We present a procedure which allows us to recover classical and nonclassical logical structures as \emph{concrete logics} associated with physical theories expressed by means of classical languages. This procedure consists in choosing, for a given theory $\mathcal T$ and classical language $\mathcal L$ expressing $\mathcal T$, an observative sublanguage $L$ of $\mathcal L$ with a notion of truth as correspondence, introducing in $L$ a derived and theory--dependent notion of \emph{C--truth} (\emph{true with certainty}), defining a \emph{physical preorder} induced by C--truth, and finally selecting a set of sentences that are \emph{verifiable} (or \emph{testable}) according to $\mathcal T$, on which a \emph{weak complementation} is induced by $\mathcal T$. The triple consisting of the set of verifiable sentences, physical order and weak complementation is then the desired concrete logic. By applying our procedure we recover a classical logic as the concrete logic associated with classical mechanics and standard quantum logic as the concrete logic associated with quantum mechanics. We also show that our alternative view of standard quantum logic, which can be constructed in a purely formal way, can be provided with a physical meaning by adopting a recent interpretation of quantum mechanics that reinterprets quantum probabilities as conditional on detection rather than absolute. Our results then show that some nonstandard logics can be obtained as mathematical structures formalizing the properties of different notions of verifiability in different physical theories. More generally, they strongly support the idea that many nonclassical logics can coexist without conflicting with classical logic (\emph{global pluralism}), for they formalize metalinguistic notions that do not coincide with the notion of truth (described by Tarski's truth theory).
\end{abstract}

\section{Introduction\label{intro}}
After the birth of modern formal logic in the XIX century many nonstandard logics were proposed, from Heyting's \emph{intuitionistic logic} \cite{h34,h56} and {\L}ukasiewicz's \emph{many valued logic} \cite{l20} to the recent \emph{relevance logic} \cite{ab75,abd92} and \emph{linear logic} \cite{g87}.

In many cases these logics are interpreted as formalizing the features of notions of truth that are alternative to the classical notion of truth as correspondence formalized by Tarski \cite{t01,t02}, and many logicians maintain that the \emph{plurality of logics} should be considered as an important achievement of the XX century, which parallels the \emph{plurality of geometries} which constitutes one of the revolutionary results of the mathematical research in the XIX century. Other logicians and philosophers instead uphold the thesis that nonstandard logics can be recovered as fragments of a suitable extension of classical logic (CL), in a unified view (\emph{global pluralism}) which restores the unity of logic and avoids many quarrels about the notion of truth \cite{h74,h78}. In this perspective, for instance, it has been proven that intuitionistic logic can be recovered as a part of a pragmatic extension of CL, intended to formalize the features of the notion of logical proof rather than the properties of an intuitionistic nonclassical notion of truth \cite{dpg95}.

In the physical realm, many nonstandard logics were propounded after the birth of quantum mechanics (QM) \cite{j74}. One only of them, however, acquired growing importance in the literature, that is, the quantum logic (QL) propounded by Birkhoff and von Neumann \cite{bvn36} as the underlying logic of QM, now often called \emph{standard} (\emph{sharp}) \emph{QL}. This proposal indeed become rather popular because it seems to spring out directly from the mathematical formalism of QM. Moreover, many scholars maintain that some known paradoxes of QM could follow from an improper use of CL for dealing with the basic notions of QM, for this theory would implicitly introduce a nonclassical notion of truth (here, briefly, \emph{quantum truth}, or \emph{Q--truth}) whose features are formalized by standard QL \cite{r98,dcgg04}. Hence a huge literature was produced on this topic, which is still flourishing nowadays.

Also in the specific case of standard QL, however, one may wonder whether a perspective of global pluralism can be adopted which allows one to recover standard QL within an extended classical framework. In particular, it has been asserted several years ago that standard QL could be seen as the mathematical structure resulting from selecting a subset of sentences that are testable according to QM in the set of all sentences of a suitable classical language, that is, as the structure formalizing the features of the metalinguistic notion of \emph{testability} in QM rather than a notion of Q--truth \cite{g92}. The aim of the present paper is to generalize and implement this view. To be precise we intend to illustrate a procedure for obtaining a \emph{concrete} (\emph{theory--dependent}) \emph{logic} associated with a physical theory $\mathcal T$ expressed by means of a classical language $\mathscr L$ with a (Tarskian) notion of truth as correspondence. This procedure consists of four steps.

(i) Consider an observational sublanguage $L$ of $\mathscr L$ and introduce a derived notion of \emph{true with certainty} on $L$ (\emph{C--truth}), defined in terms of classical truth but depending on $\mathcal T$.

(ii) Define a \emph{physical preorder} $\prec$ on $L$, induced by the notion of C--truth.

(iii) Introduce a notion of \emph{verifiability} in $L$ by selecting a subset $\phi_{V}$ of sentences of $L$ that are \emph{verifiable}, or \emph{testable}, according to $\mathcal T$.

(iv) Define a \emph{weak complementation} ${}^{\perp}$ induced by $\mathcal T$ on $(\phi_{V}, \prec)$.

The structure $(\phi_V, \prec, {}^{\perp})$ is then the required concrete logic.

The above procedure is conceptually and philosophically relevant because it can explain the origins of some important nonstandard logics. Indeed one can obtain different concrete logics by changing some of the three basic elements of the procedure, that is, $\mathcal T$, $L$ and the notion of verification. 

We illustrate our procedure in Secs. \ref{linguaggio} and \ref{physics} by introducing a very simple observational (pretheoretical) language $L(x)$ which is suitable for expressing basic notions and relations in a wide class of physical theories. Then we consider two fundamental theories of modern physics, \emph{i.e.}, classical mechanics (CM) and QM. We obtain in Sec. \ref{fisicaclassica} a classical logical structure as the concrete logic associated with CM if all sentences of $L(x)$ are considered verifiable (in principle), as usual in CM. But we also prove by means of an example in Sec. \ref{nonBoolean} that one can obtain a concrete logic that exhibits a non--Boolean lattice structure in a macroscopic domain in which CM holds if one selects a suitable language $L$ and subset of sentences of $L$ that are considered verifiable. Furthermore, we show in Sec. \ref{meccanicaquantistica} that, if $\mathcal T$ is QM, our procedure allows one to recover standard QL as the concrete logic associated with QM (up to an equivalence relation) if a standard notion of verifiability according to QM is adopted. Finally we compare in the last part of the paper (Sec. \ref{quantum_logic}) our procedure for obtaining standard QL with the orthodox procedure that introduces standard QL without making reference to classical truth \cite{r98,dcgg04}, and show that the latter implies adopting implicitly a \emph{verificationist} (hence theory dependent) definition of truth, while truth and verification are carefully distinguished in our approach.

Let us add now some comments and remarks on the results resumed above.

Firstly, we observe that our approach does not conflict with \emph{logical localism} \cite{dc00}, for it entails that many nonstandard logics can be obtained, depending both on the theory $\mathcal T$ and on the notion of verifiability that is considered. Nevertheless it also supports global pluralism, because it implies that different concrete logics can coexist without contradicting classical logic. 

Secondly, we note that our procedure has to cope with an important objection. Indeed, a quantum physicist could argue that it recovers standard QL in a classical framework in a  purely formal way because QM cannot be expressed by a classical language. More specifically, he would observe that we associate a set with every property of a physical system (the \emph{extension} of the property) whose elements are interpreted as individual examples of the system possessing the property, while such a set cannot be defined according to the orthodox interpretation of QM. Physical properties are in fact \emph{nonobjective} according to this interpretation, which means that the set of individual examples of the physical system in a given state that display a given property whenever this property is measured depends on the set of measurements that are actually performed and is not prefixed (hence one cannot say that these examples ``possess'' the property before the measurement). To overcome this objection, we preliminarily note that our formal derivation of standard QL is interesting anyway, for it is usually maintained that any derivation of this kind is impossible. More important, the objection vanishes if one accepts the point of view of some alternative interpretations of QM in which objectivity of physical properties is recovered. In particular, an extensive criticism of the theorems that aim to prove the \emph{contextuality} and the \emph{nonlocality} of QM (mainly the Bell--Kochen--Specker \cite{b66,ks67} and the Bell \cite{b64} theorems), hence nonobjectivity of physical properties, has been carried out by some authors who have shown that the proofs of these theorems implicitly introduce an epistemological assumption which is problematic from the point of view of QM. If this assumption is relaxed, the proofs cannot be completed \cite{gs96a,gs96b,gp04}. Basing on this conclusion, a new theoretical proposal called \emph{extended semantic realism} (ESR) \emph{model} has been recently worked out. The ESR model reinterprets quantum probabilities as conditional on detection rather than absolute and embodies the mathematical formalism of QM into a broader formalism which admits a local and noncontextual physical interpretation \cite{gs09b,gs08,gs10a}, recovering objectivity of physical properties. If this new perspective is accepted, our present deduction of standard QL is not purely formal because its basic elements are provided with a physical interpretation. 

Thirdly, we note that we recover standard QL in this paper by introducing semantic restrictions induced by QM on a classical language and then defining new derived quantum connectives whose semantic interpretation is provided by QM. It is then possible to show that an equivalent approach to standard QL can be constructed by pragmatically extending a classical propositional logic by means of pragmatic connectives, following \cite{dpg95}, and then interpreting these connectives in terms of the quantum notion of \emph{empirical proof} (which is equivalent to the notion of verification according to QM that is introduced in this paper).

We conclude by noticing that the elementary observational sublanguage $L(x)$ introduced in this paper takes into account only pure states and physical properties (exact effects) and does not contain logical quantifiers, to avoid clouding the conceptual aspects of our approach with technical complications. In principle, however, it can be extended to take into account mixtures and effects and express general empirical laws.

\section{The language $\bf{L(x)}$} \label{linguaggio}
Let $\mathcal T$ be a physical theory in which the following notions are introduced \cite{bc81,l83}.

\noindent
\emph{Physical system}. 

\noindent
\emph{Physical property} of a physical system, operationally defined as a class of dichotomic registering devices that are physically equivalent according to $\mathcal T$.

\noindent 
\emph{Physical state} of a physical system, operationally defined as a class of preparing devices that are physically equivalent according to $\mathcal T$.

\noindent
\emph{Physical object}, operationally defined as an individual example of a physical system, prepared by means of a preparing device. 

We denote by $L(x)$ a formal language intended to express basic notions and relations in $\mathcal T$, constructed as follows.

\vspace{.2cm}
\noindent
\textbf{Def. 2.1.} \emph{The} alphabet \emph{of $L(x)$ consists of the following elements.} \\
\emph{Two disjoint sets of monadic predicates, ${\mathcal E}= \{ E, F, \ldots \}$ (intended interpretation: physical properties) and ${\mathscr S}= \{ S, T, \ldots \}$ (intended interpretation: pure states).} \\
\emph{Individual variable $x$.} \\
\emph{Connectives $\lnot, \land, \lor, \rightarrow $.} \\
\emph{Parentheses $(\, , \,)$.}

\vspace{.2cm}
\noindent
\textbf{Def. 2.2.} \emph{The set $\psi(x)$ of all} well formed formulas \emph{(}wffs\emph{) of $L(x)$ is the set obtained by applying recursively standard} formation rules \emph{in CL (to be precise, for every $A \in {\mathcal E} \cup {\mathscr S}$, $A(x) \in \psi(x)$; for every $\alpha(x) \in \psi(x)$, $\lnot\alpha(x) \in \psi(x)$; for every $\alpha(x), \beta(x) \in \psi(x)$, $\alpha(x) \land \beta(x) \in \psi(x)$, $\alpha(x) \lor \beta(x) \in \psi(x)$, $\alpha(x) \rightarrow \beta(x) \in \psi(x)$). Furthermore, we put ${\mathcal E}(x)= \{ E(x) \ | \ E \in {\mathcal E} \}$, ${\mathscr S}(x)=\{ S(x) \ | \ S \in {\mathscr S} \}$ and call ${\mathcal E}(x) \cup {\mathscr S}(x)$ the set of all} atomic \emph{wffs of $\psi(x)$.}

\vspace{.2cm}
\noindent
\textbf{Def. 2.3.} \emph{The} semantics \emph{of $L(x)$ consists of the following elements.} \\
\emph{Universe ${\mathcal U}$ (intended interpretation: set of physical objects).} \\
\emph{Injective mapping $ext: A \in {\mathcal E} \cup {\mathscr S} \longmapsto ext (A) \in {\mathscr P}({\mathcal U})$ (power set of $\mathcal U$).} \\
\emph{Boolean lattice $\langle ext ({\mathcal E} \cup {\mathscr S}) \rangle$ generated by $ext({\mathcal E} \cup {\mathscr S})$ via ${}^{c}, \cap, \cup$.} \\
\emph{Recursive definition of the surjective mapping (still called  \emph{$ext$} by abuse of language)}
\begin{center}
\emph{$ext: \alpha(x) \in \psi(x) \longmapsto ext (\alpha(x)) \in \langle ext ({\mathcal E} \cup {\mathscr S}) \rangle$}
\end{center}
\emph{(to be precise, for every $E \in {\mathcal E}$, $ext (E(x))= ext (E)$; for every $S \in {\mathscr S}$, $ext (S(x))= ext (S)$; for every $\alpha(x) \in \psi(x)$, $ext (\lnot \alpha(x))={\mathcal U}\setminus ext(\alpha(x))=(ext (\alpha(x)))^{c}$; for every $\alpha(x), \beta(x)\in \psi(x)$, $ext(\alpha(x) \land \beta(x))=ext (\alpha(x)) \cap ext (\beta(x))$, $ext(\alpha(x) \lor \beta(x))=ext (\alpha(x)) \cup ext (\beta(x))$, $ext(\alpha(x) \rightarrow \beta(x)) = (ext(\alpha(x)))^{c} \cup ext(\beta(x))$).} \\
\emph{Interpretation of the variable $\sigma: x \in \{ x \} \longmapsto \sigma(x) \in {\mathcal U}$.} \\
\emph{Set $\Sigma$ of all interpretations of the variable.} \\
\emph{For every $\sigma \in \Sigma$, truth assignment}
\begin{center}
\emph{$v_{\sigma}: \alpha(x) \in \psi(x) \longmapsto v_{\sigma}(\alpha(x))\in \{ T, F \}$}
\end{center}
\emph{such that $v_{\sigma}(\alpha(x))=T (F)$ iff $\sigma(x) \in ext (\alpha(x))$ ($\sigma(x) \in (ext (\alpha(x)))^{c}$).} \\
\emph{Definition of truth (falsity): for every $\alpha(x) \in \psi(x)$, $\alpha(x)$ is} true \emph{(}false\emph{) in $\sigma$ iff $v_{\sigma}(\alpha(x))=T (F)$.}

\vspace{.2cm}
\noindent
\textbf{Def. 2.4.} \emph{The binary relations of} logical preorder \emph{$\le$ and} logical equivalence \emph{$\equiv$ on $\psi(x)$ are defined by setting, for every $\alpha(x),\beta(x)\in \psi(x)$,} 
\begin{center}
\emph{$\alpha(x) \le \beta(x)$ iff (for every $\sigma \in \Sigma$, $v_{\sigma}(\alpha(x))=T$ implies $v_{\sigma}(\beta(x))=T$)} 
\end{center}
\emph{and} 
\begin{center}
\emph{$\alpha(x) \equiv \beta(x)$ iff (for every $\sigma \in \Sigma$, $v_{\sigma}(\alpha(x))=T$ iff $v_{\sigma}(\beta(x))=T$).}
\end{center}

One can then prove some statements that are useful to compare the logical notions introduced in this section with the physical notions introduced in the following sections.

\vspace{.2cm}
\noindent
\textbf{Prop. 2.1.} \emph{(i) $ext (\psi(x))= \langle ext ({\mathcal E} \cup {\mathscr S}) \rangle$, hence $(ext (\psi(x)), \cap, \cup, {}^{c})$ is a Boolean algebra; equivalently, $(ext (\psi(x)), \subset, {}^{c})$ is a Boolean lattice.}

\emph{(ii) For every $\alpha(x),\beta(x)\in \psi(x)$,}
\begin{center}
\emph{$\alpha(x) \le \beta(x)$ iff $ext (\alpha(x)) \subset ext (\beta(x))$}
\end{center}
\emph{and}
\begin{center}
\emph{$\alpha(x) \equiv \beta(x)$ iff ($\alpha(x) \le \beta(x)$ and $\beta(x) \le \alpha(x)$) iff $ext (\alpha(x)) = ext (\beta(x))$.} 
\end{center}

\emph{(iii) The equivalence relation $\equiv$ is compatible with $\lnot,\land,\lor$ and $\rightarrow$ (that is, for every $\alpha(x), \beta(x),\gamma(x), \delta(x)\in \psi(x)$, $\alpha(x)\equiv \beta(x)$ implies $\lnot \alpha(x)\equiv \lnot \beta(x)$, ($\alpha(x)\equiv\gamma(x)$ and $\beta(x)\equiv\delta(x)$) implies $\alpha(x)\land\beta(x)\equiv\gamma(x)\land\delta(x)$, etc.).}

\emph{(iv) The structure $(\psi(x)/_{\equiv},\land',\lor',\lnot')$ (where $\land',\lor',\lnot'$ denote the operations canonically induced on $\psi(x)/_{\equiv}$ by $\land,\lor,\lnot$, respectively) is a Boolean algebra isomorphic to $(ext (\psi(x)), \cap,\cup, {}^{c})$ (\emph{Lindenbaum--Tarski algebra} of $L(x)$); equivalently, $(\psi(x)/_{\equiv}, \le', \lnot')$ (where $\le'$ is the order canonically induced on $\psi(x)/_{\equiv}$ by the preorder $\le$ defined on $\psi(x)$) is a Boolean lattice isomorphic to $(ext (\psi(x)), \subset, {}^{c})$.}

\section{General physical conditions on $\bf{L(x)}$\label{physics}}
The notions introduced in this section hold for any physical theory $\mathcal T$ of the kind considered at the beginning of Sec. \ref{linguaggio}, are suggested by the intended interpretation of $L(x)$ and are partially illustrated by the drawing in Fig. 1. For the sake of brevity we shall understand the word ``pure'' when referring to states in the following.

\vspace{.2cm}
\noindent
\textbf{Axiom P.} \emph{$\{ext (S) \ | \ S \in {\mathscr S} \}$ is a partition of $\mathcal U$.}

\vspace{.2cm}
\noindent
\emph{Physical justification}. States are defined as equivalence classes of preparations (Sec. \ref{linguaggio}) and every physical object in $\mathcal U$ is prepared by one and only one preparation, hence it belongs to one and only one extension of a state.

The following statement is then an immediate consequence of Axiom P.

\vspace{.2cm}
\noindent
\textbf{Prop. 3.1.} \emph{For every $\sigma \in \Sigma$, there is one and only one state $S_{\sigma} \in {\mathscr S}$ such that $\sigma(x) \in ext (S_{\sigma})$  (equivalently, $v_{\sigma}(S_{\sigma}(x))=T$).}
\begin{figure}
\begin{center}
\includegraphics[scale=0.7]{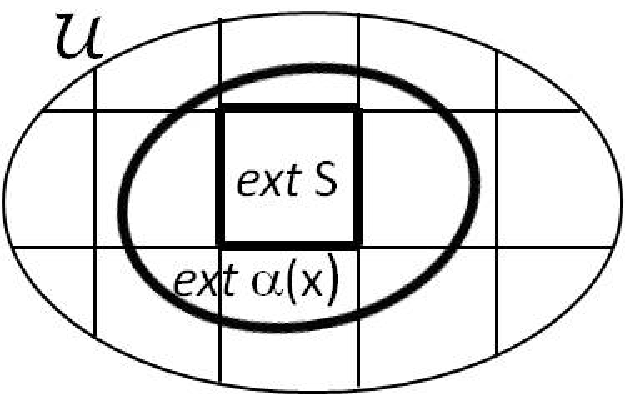}
\end{center}
\emph{Fig. 1.}
\end{figure}

\vspace{.2cm}
\noindent
In a physical theory $\mathcal T$ of the kind considered in Sec. \ref{linguaggio} it is customary to associate every physical property $E$ with a set of states ${\mathscr S}_{E}$ such that, if a physical object $x$ is prepared by a preparing device belonging to the state $S \in {\mathscr S}_{E}$, then $x$ displays the property $E$ with certainty whenever a measurement of $E$ is performed.  The following definition generalizes this idea.

\vspace{.2cm}
\noindent
\textbf{Def. 3.1.} \emph{We denote by $\phi(x)$ the subset of wffs of $\psi(x)$ which are constructed by using only atomic wffs in ${\mathcal E}(x)$, hence do not contain predicates denoting states. Then, for every $\alpha(x) \in \phi(x)$, we call} physical proposition \emph{associated with $\alpha(x)$ the set of states}
\begin{center}
\emph{$p_\alpha= \{ S \in {\mathscr S} \ | \ ext (S) \subset ext (\alpha(x)) \}$}
\end{center}
\emph{and denote by $\mathcal P$ the set of all physical propositions associated with wffs of $\phi(x)$.}

\vspace{.2cm}
\noindent
The proof of the following statement is then straightforward.

\vspace{.2cm}
\noindent
\textbf{Prop. 3.2} \emph{For every $\alpha(x) \in \phi(x)$,}
\begin{center}
\emph{$p_\alpha= \{ S \in {\mathscr S} \ | \ \textrm{for every} \ \sigma \in \Sigma, \ v_{\sigma}(S(x) \rightarrow \alpha(x))=T \}$.}
\end{center}

\vspace{.2cm}
\noindent
By using the notion of physical proposition one can define a notion of \emph{C--truth} on $\phi(x)$ (which depends on $\mathcal T$ because, for every $\alpha(x) \in \phi(x)$, the set of states in the physical proposition $p_{\alpha}$ depends on $\mathcal T$).

\vspace{.2cm}
\noindent
\textbf{Def. 3.2.} \emph{For every $\alpha(x) \in \phi(x)$, let us denote by $p_{\lnot\alpha}$ the physical proposition associated with $\lnot\alpha(x)$. Then, for every $S \in {\mathscr S}$, we say that $\alpha(x)$ is} C--true \emph{(}certainly true\emph{) in $S$ iff $S \in p_{\alpha}$,} C--false \emph{(}certainly false\emph{) in $S$ iff $S \in p_{\lnot\alpha}$. Furthermore, we say that $\alpha(x)$ has no C--truth value (or that it is} C--indeterminate\emph{) iff $S \notin p_{\alpha} \cup p_{\lnot\alpha}$.}

\vspace{.2cm}
\noindent
One can then easily prove the following statement.

\vspace{.2cm}
\noindent
\textbf{Prop. 3.3.} \emph{For every $S \in {\mathscr S}$ and $\alpha(x) \in \phi(x)$,}
\begin{center}
\emph{$\alpha(x)$ is C--true in $S$ iff $ext (S) \subset ext (\alpha(x))$ iff (for every $\sigma \in \Sigma$, $S(x)$ is true in $\sigma$ implies $\alpha(x)$ is true in $\sigma$) iff for every $\sigma \in \Sigma$, $v_{\sigma}(S(x)\rightarrow \alpha(x))=T$,}
\end{center}
\emph{and}
\begin{center}
\emph{$\alpha(x)$ is C--false in $S$ iff $ext (S) \subset (ext (\alpha(x)))^{c}$ iff (for every $\sigma \in \Sigma$, $S(x)$ is true in $\sigma$ implies $\lnot\alpha(x)$ is true in $\sigma$) iff for every $\sigma \in \Sigma$, $v_{\sigma}(S(x)\rightarrow \lnot\alpha(x))=T$.}
\end{center}

\vspace{.2cm}
\noindent
By using the notion of C--truth one can introduce new binary relations on $\phi(x)$, as follows.

\vspace{.2cm}
\noindent
\textbf{Def. 3.3.} \emph{We call} physical preorder \emph{$\prec$ and} physical equivalence \emph{$\approx$ the binary relations defined on $\phi(x)$ by setting, for every $\alpha(x), \beta(x) \in \phi(x)$,}
\begin{center}
\emph{$\alpha(x) \prec \beta(x)$ iff (for every $S \in {\mathscr S}$, $\alpha(x)$ is C--true in $S$ implies $\beta(x)$ is C--true in $S$)}
\end{center}
\emph{and}
\begin{center}
\emph{$\alpha(x) \approx \beta(x)$ iff (for every $S \in {\mathscr S}$, $\alpha(x)$ is C--true in $S$ iff $\beta(x)$ is C--true in $S$).}
\end{center}
\emph{Furthermore, we denote by $\prec'$ the order canonically induced on $\phi(x)/_{\approx}$ by the preorder $\prec$ defined on $\phi(x)$.}

\vspace{.2cm}
\noindent
The proofs of the following statements are then straightforward.

\vspace{.2cm}
\noindent
\textbf{Prop. 3.4.} \emph{(i) For every $\alpha(x), \beta(x) \in \phi(x)$,}
\begin{center}
\emph{$\alpha(x) \prec \beta(x)$ iff $p_{\alpha} \subset p_{\beta}$}
\end{center}
\emph{and}
\begin{center}
\emph{$\alpha(x) \approx \beta(x)$ iff $p_{\alpha} = p_{\beta}$ iff ($\alpha(x) \prec \beta(x)$ and $\beta(x) \prec \alpha(x)$).}
\end{center}

\emph{(ii) $(\phi(x)/_{\approx}, \prec')$ is isomorphic to $({\mathcal P}, \subset)$.}

\emph{(iii) Let $\alpha(x), \beta(x) \in \phi(x)$. Then}
\begin{center}
\emph{$\alpha(x) \le \beta(x)$ \ implies \ $\alpha(x) \prec \beta(x)$}
\end{center}
\emph{and}
\begin{center}
\emph{$\alpha(x) \equiv \beta(x)$ \ implies \ $\alpha(x) \approx \beta(x)$.}
\end{center}

\vspace{.2cm}
\noindent
We observe now that the intended interpretation of $L(x)$ implies that a wff $\alpha(x) \in \phi(x)$ can be verified whenever a dichotomic registering device exists which, for every $\sigma \in \Sigma$, may perform a measurement on $\sigma(x)$ specifying whether $\alpha(x)$ is true or false in $\sigma$. But a dichotomic registering device of this kind characterizes a physical property $E$ in the theory $\mathcal T$ (Sec. \ref{linguaggio}), hence we are led to introduce the following definition.

\vspace{.2cm}
\noindent
\textbf{Def. 3.4.} \emph{We call} set of verifiable\emph{, or} testable\emph{,} wffs \emph{of $L(x)$ the subset}
\begin{center}
\emph{$\phi_{V}(x)=\{ \alpha(x) \in \phi(x) \ | \ \exists E_{\alpha} \in {\mathcal E}: \ \alpha(x) \equiv E_{\alpha}(x) \} \subset \phi(x)$}
\end{center}
\emph{and call} set of verifiable\emph{, or} testable\emph{, propositions of $L(x)$ the subset}
\begin{center}
\emph{${\mathcal P}_{V}=\{  p_{\alpha}\in {\mathcal P} \ | \ \alpha(x) \in \phi_{V}(x) \} \subset {\mathcal P}$.}
\end{center}

\vspace{.2cm}
\noindent
Basing on Def. 3.4 one can prove at once the following statements.

\vspace{.2cm}
\noindent
\textbf{Prop. 3.5.}
\emph{(i) $\phi_{V}(x)=\{ \alpha(x) \in \phi(x) \ | \ \exists E_{\alpha} \in {\mathcal E}: \ ext(\alpha(x))= ext(E_{\alpha}(x)) \}$.}

\emph{(ii) The following order structures and isomorphisms can be introduced basing on the logical and physical orders.}

\emph{$(\phi(x), \le)$.} 

\emph{$(\phi(x)/_{\equiv}, \le')$, isomorphic to $(ext (\phi(x)), \subset)$.} 

\emph{$(\phi_{V}(x), \le)$.} 

\emph{$(\phi_{V}(x)/_{\equiv}, \le')$, isomorphic to $(ext ({\mathcal E}), \subset)$.}

\emph{$(\phi(x), \prec)$.} 

\emph{$(\phi(x)/_{\approx}, \prec')$, isomorphic to $({\mathcal P}, \subset)$.} 

\emph{$(\phi_{V}(x), \prec)$.}

\emph{$(\phi_{V}(x)/_{\approx}, \prec')$, isomorphic to $({\mathcal P}_{V}, \subset)$.}

\vspace{.2cm}
\noindent
The crucial notion of logic associated with a physical theory $\mathcal T$, distinguished from (but related to) the logical structure of the language by means of which $\mathcal T$ is expressed, can now be introduced by means of the following definition.

\vspace{.2cm}
\noindent
\textbf{Def. 3.5.} \emph{Whenever the theory $\mathcal T$ induces a} weak complementation \emph{${}^{\perp}$ on $(\phi_{V}(x), \prec)$ (\emph{i.e.}, a mapping ${}^{\perp}: \alpha(x) \in \phi_{V}(x) \longmapsto (\alpha(x))^{\perp} \in \phi_{V}(x)$ such that $(\alpha(x))^{\perp\perp} \approx \alpha(x)$ and $\alpha(x) \prec\beta(x)$ implies $(\beta(x))^{\perp} \prec (\alpha(x))^{\perp}$) we say that the set $(\phi_{V}(x), \prec, {}^{\perp})$ is the} concrete logic of $\mathcal T$.

\vspace{.2cm}
\noindent
We stress that the set $\phi_V(x)$ is selected by adopting a standard notion of verification that holds both in CM and in QM. Of course, different choices of $\phi_V(x)$ may lead one to associate different concrete logics with $\mathcal T$. 

\begin{figure}
\begin{center}
\includegraphics[scale=0.7]{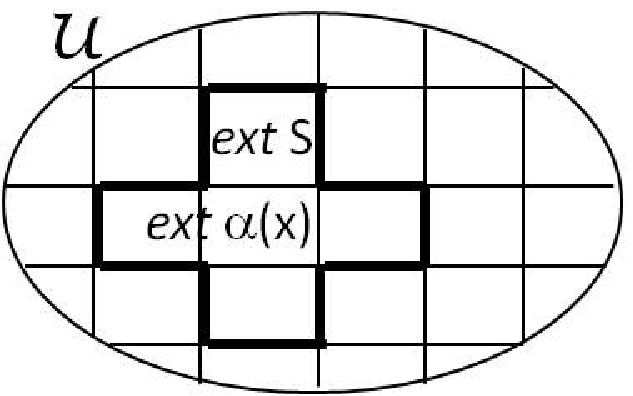}
\end{center}
\emph{Fig. 2.}
\end{figure}
\section{Classical mechanics and classical logic\label{fisicaclassica}}
We intend to show in this section that the concrete logic of CM within the general classical approach sketched in Secs. \ref{linguaggio} and \ref{physics} has the structure of a classical logic. To this end we preliminarily state the following fundamental axiom of CM, which establishes the mathematical representation of states and physical properties in CM.

\vspace{.2cm}
\noindent
\textbf{Axiom CM1.} \emph{Every physical system is represented in CM by a triple $({\mathcal F}, \varphi, \chi)$, where ${\mathcal F}, \varphi$ and $\chi$ are defined as follows.}

\emph{${\mathcal F}$ is a phase space associated with the physical system $\Omega$.}

\emph{$\varphi: S \in {\mathscr S} \longmapsto \varphi(S) \in {\mathcal F}$.}

\emph{$\chi: E \in {\mathcal E} \longmapsto \chi(E) \in {\mathscr P}({\mathcal F})$} 

\noindent
\emph{where ${\mathscr P}({\mathcal F})$ is the power set of ${\mathcal F}$.}

\emph{The mappings $\varphi$ and $\chi$ are bijective.}

\vspace{.2cm}

%\begin{figure}
%\begin{center}
%\includegraphics[scale=0.6]{fig3}
%\end{center}
%\emph{Fig. 3.}
%\end{figure}
We introduce now two axioms that can be justified in CM on the basis of the intended interpretation provided in Sec. \ref{linguaggio}. The first of them relates the mathematical representation in Axiom CM1 with the set--theoretical representations of states and physical properties provided in Sec. \ref{linguaggio}.

\vspace{.2cm}
\noindent
\textbf{Axiom CM2.} \emph{For every $E \in {\mathcal E}$ and $S \in {\mathscr S}$,}
\begin{center}
\emph{$\varphi(S) \in \chi(E)$ iff $ext (S) \subset ext (E)$,} \\
\emph{$\varphi(S) \in {\mathcal F} \setminus \chi(E)$ iff $ext (S) \subset {\mathcal U} \setminus ext (E)$.}
\end{center}

\vspace{.2cm}
\noindent
\emph{Physical justification}. All physical objects in a given state (that is, prepared by preparations that are physically equivalent in the sense established by CM) either possess or do not possess a given physical property according to CM.

By using Axiom CM2 one can easily prove the following statements (which are partially illustrated by the drawing in Fig. 2).

\vspace{.2cm}
\noindent
\textbf{Prop. 4.1.} \emph{(i) For every $E \in {\mathcal E}$ and $S \in {\mathscr S}$, either $ \ ext (S) \subset ext (E)$ or $ext (S) \subset {\mathcal U} \setminus ext (E)=(ext(E))^{c}$.}

\emph{(ii) For every $\alpha(x) \in \phi(x)$ and $S \in {\mathscr S}$, either $ext(S) \subset ext(\alpha(x))$ or $ext(S) \subset {\mathcal U} \setminus ext(\alpha(x))=(ext(\alpha(x)))^{c}$ (hence $ext (\alpha(x))=\cup \{ext (S) \ | \ ext (S) \subset ext (\alpha(x)) \}$.}

\emph{(iii) For every $\sigma \in \Sigma$ and $\alpha(x) \in \phi(x)$,}
\begin{center}
\emph{$\alpha(x)$ is true (false) in $\sigma$ iff $\alpha(x)$ is C--true (C--false) in $S_{\sigma}$ \\
 iff $ext (S_{\sigma}) \subset ext (\alpha(x))$ ($ext (S_{\sigma}) \subset (ext (\alpha(x)))^{c}$) iff $S_{\sigma}\in p_{\alpha}$ ($S_{\sigma}\in {\mathscr S} \setminus p_{\alpha}$).}
\end{center}

\emph{(iv) For every $\alpha(x),\beta(x) \in \phi(x)$,}
\begin{center}
\emph{$\alpha(x) \le \beta(x)$ \ iff \ $\alpha(x) \prec \beta(x)$}
\end{center}
\emph{and}
\begin{center}
\emph{$\alpha(x) \equiv \beta(x)$ \ iff \ $\alpha(x) \approx \beta(x)$.}
\end{center}

\vspace{.2cm}
\noindent
It is important to note that Prop. 4.1, (iii), shows that truth and C--truth coincide in CM. Indeed this coincidence explains why no distinction between the two notions of truth is made in CM.

The last axiom then characterizes the notion of verification in CM.

\vspace{.2cm}
\noindent
\textbf{Axiom CM3.} \emph{The set $\phi_{V}(x)$ of all verifiable wffs of $\phi(x)$ coincides with $\phi(x)$.}

\vspace{.2cm}
\noindent
\emph{Physical justification}. All physical statements about physical objects are testable, in principle, according to CM.

By using Axiom CM3 one can easily prove the following statements.

\vspace{.2cm}
\noindent
\textbf{Prop. 4.2.} \emph{(i) For every $\sigma \in \Sigma$ and $\alpha(x) \in \phi(x)$,}
\begin{center}
\emph{$\alpha(x)$ is true (false) in $\sigma$ iff $E_{\alpha}(x)$ is true (false) in $\sigma$ iff $ext(S_{\sigma}) \subset ext(E_{\alpha})$  ($ext(S_{\sigma}) \subset (ext(E_{\alpha}))^{c}$).}
\end{center}

\emph{(ii) For every $\alpha(x), \beta(x) \in \phi(x)$,}
\begin{center}
\emph{$\alpha(x) \le \beta(x)$ iff $E_{\alpha}(x) \le E_{\beta}(x)$,}
\end{center}
\emph{and}
\begin{center}
\emph{$\alpha(x) \equiv \beta(x)$ iff $E_{\alpha}(x) \equiv E_{\beta}(x)$.}
\end{center}

\emph{(iii) $(\phi(x), \le)$, $(\phi_{V}(x), \le)$, $(\phi(x), \prec)$ and $(\phi_{V}(x), \prec)$ can be identified.}

\emph{(iv) $(\phi(x)/_{\equiv}, \le')$, $(\phi_{V}(x)/_{\equiv}, \le')$, $(\phi(x)/_{\approx}, \prec')$ and $(\phi_{V}(x)/_{\approx}, \prec')$ can be identified. Furthermore, they are isomorphic to $(ext (\phi(x)), \subset)$, $(ext ({\mathcal E}), \subset)$, $({\mathcal P}, \subset)$ and $({\mathcal P}_{V}, \subset)$.}

\emph{(v) The mapping}
\begin{center}
\emph{${}^{\perp}: \alpha(x) \in \phi_{V}(x) \longmapsto \lnot\alpha(x) \in \phi_{V}(x)$}
\end{center}
\emph{is a weak complementation on $(\phi_{V}(x), \prec)$.}

\emph{(vi) The structure $(\phi_{V}(x), \prec, {}^{\perp})$ is the concrete logic of CM and \mbox{$(\phi_{V}(x)/_{\approx}, \prec', {}^{\perp'})$} (where ${}^{\perp'}$ is the complementation canonically induced by ${}^{\perp}$ on $\phi_{V}(x)/_{\equiv}$) is a Boolean lattice that can be identified with the Boolean algebra $(\phi_V(x)/_{\equiv},\land',\lor',\lnot')$.}

\vspace{.2cm}
\noindent
The result in Prop. 4.2, (vi), explains from our present standpoint the common statement in the literature that ``the logic of a classical physical system is classical logic''. We stress, however, that this statement follows from Axioms CM2 and CM3 and could not be proven if these axioms should not hold.

For the sake of completeness we add the following statements, which establish further links between the mathematical representation in Axiom CM1 and the set--theoretical representations of states and physical properties provided in Sec. \ref{linguaggio}.

\vspace{.2cm}
\noindent
\textbf{Prop. 4.3.} \emph{(i) For every $E,F \in {\mathcal E}$,}
\begin{center}
\emph{$\chi(E) \subset \chi(F)$ iff $ext (E) \subset ext (F)$.}
\end{center}
\emph{(ii) For every $\sigma \in \Sigma$ and $\alpha(x) \in \phi(x)$,}
\begin{center}
\emph{$\alpha(x)$ is true (false) in $\sigma$ iff $\varphi(S_{\sigma})\in \chi(E_{\alpha})$ ($\varphi(S_{\sigma})\in {\mathcal F} \setminus \chi(E_{\alpha})$).}
\end{center}
\emph{(iii) For every $\alpha(x) \in \phi(x)$,}
\begin{center}
\emph{$p_{\alpha}= \{ S \in {\mathscr S} \ | \ \varphi(S) \in \chi(E_{\alpha}) \}$.}
\end{center}

\section{Non--Boolean structures in classical mechanics\label{nonBoolean}}
Prop. 4.2, (vi), states that $(\phi_{V}(x)/_{\approx}, \prec', {}^{\perp'})$ is a Boolean lattice in CM which can be identified with the Boolean algebra $(\phi(x)/_{\equiv},\land',\lor',\lnot')$. It is then important to observe that non--Boolean algebras can be obtained in CM if the testability criteria are suitably restricted. For instance, let us assume that not all wffs in $\phi_{V}(x)$, which are testable in principle, are testable in practice, and let us introduce the subset $\phi_{V}^{P}(x)\subset \phi_{V}(x)$ of wffs of $L(x)$ that are \emph{practically testable}. In this case we can consider the preordered set $(\phi_{V}^{P}(x), \le)$ and the quotient set $(\phi_{V}^{P}(x)/_{\equiv}, \le')$. Yet, the latter generally is not a Boolean lattice, and it can be equipped with different algebraic structures by choosing $\phi_{V}^{P}(x)$ in different ways \cite{g92}. %Typical examples of implicit applications of this procedure are provided by Aerts' quantum machine \cite{a99a,a99b} and Kirkpatrick's playing cards \cite{k03,k04}. In the former case the lattice $(\phi_{V}^{P}(x)/_{\equiv}, \le')$ is isomorphic to the standard QL of a spin--$\frac{1}{2}$ physical system (which is sufficient to show that standard QL does not characterize QM, as commonly maintained). In the latter case $(\phi_{V}^{P}(x)/_{\equiv}, \le')$ also exhibits basic properties of standard QL.

It is also interesting to note that there are examples in the literature of physical systems in which quantum structures are obtained in a macroscopic domain in which CM holds. These examples are relevant because they falsify the widespread belief that QL characterizes QM, hence indirectly support our position in this paper. Let us therefore present briefly one of them, that is, Aerts' \emph{ quantum machine} \cite{a88,a91,a95,a98}.

Aerts writes in \cite{a98}:

\begin{quote}
``The machine that we consider consists of a physical entity $S$ that is a point particle $P$ that can move on the surface of a sphere, denoted $surf$, with center $O$ and radius 1. The unit--vector $v$ where the particle is located on $surf$ represents the state $p_v$ of the particle (\ldots). For each point $u \in surf$, we introduce the following measurement $e_u$. We consider the diametrically opposite point $-u$, and install a piece of elastic of length 2, such that it is fixed with one of its end--points in $u$ and the other end--point in $-u$. Once the elastic is installed, the particle $P$ falls from its original place $v$ orthogonally onto the elastic, and sticks on it (\ldots). Then the elastic breaks and the particle $P$, attached to one of the two pieces of the elastic (\ldots), moves to one of the two end--points $u$ or $-u$ (\ldots). Depending on whether the particle $P$ arrives in $u$ (\ldots) or in $-u$, we give the outcome $o_1^u$ or $o_2^u$ to $e_u$. We can easily calculate the probabilities corresponding to the two possible outcomes. Therefore we remark that the particle $P$ arrives in $u$ when the elastic breaks in a point of the interval $L_1$ (which is the length of the piece of the elastic between $-u$ and the point where the particle has arrived, or $1+\cos \theta$), [where $\theta$ is the angle between $u$ and $v$] and arrives in $-u$ when it breaks in a point of the interval $L_2$ ($L_2 = L-L_1 = 2-L_1$). We make the hypothesis that the elastic breaks uniformly, which means that the probability that the particle, being in state $p_v$, arrives in $u$, is given by the length of $L_1$ divided by the length of the total elastic (which is 2). The probability that the particle in state $p_v$ arrives in $-u$ is the length of $L_2$ (which is $1-\cos \theta$) divided by the length of the total elastic. If we denote these probabilities respectively by $P(o_1^u, p_v)$ and $P(o_2^u, p_v)$ we have:
\begin{eqnarray}
P(o_1^u, p_v)=\frac{1+\cos \theta}{2}=\cos^2\frac{\theta}{2} \nonumber \\
P(o_2^u, p_v)=\frac{1-\cos \theta}{2}=\sin^2\frac{\theta}{2} \nonumber
\end{eqnarray}
These transition probabilities are the same as the ones related to the outcomes of a Stern--Gerlach spin measurement on a spin $\frac{1}{2}$ quantum particle, \ldots''.
\end{quote}

Therefore Aerts concludes:
\begin{quote}
``We can easily see now the two aspects in this quantum machine that we have identified in the hidden measurement approach to give rise to the quantum structure. The state of the particle $P$ is effectively changed by the measuring apparatus ($p_v$ changes to $p_u$ or to $p_{-u}$ under the influence of the measuring process), which identifies the first aspect, and there is a lack of knowledge on the interaction between the measuring apparatus and the particle, namely the lack of knowledge of were exactly the elastic will break, which identifies the second aspect. We can also easily understand now what is meant by the term `hidden measurements'. Each time the elastic breaks in one specific point $\lambda$, we could identify the measurement process that is carried out afterwards as a hidden measurement $e^\lambda_u$. The measurement $e_u$ is then a classical mixture of the collection of all measurement $e^\lambda_u$: namely $e_u$ consists of choosing at random one of the $e^\lambda_u$ and performing this chosen $e^\lambda_u$.''
\end{quote}

Let us qualitatively discuss the above example from the point of view proposed in the present paper. It follows from Aerts' description that the physical system that is considered is a classical particle $P$ of which a complete deterministic description is possible, in principle, in CM. %Hence our general treatment in Secs. \ref{linguaggio}--\ref{fisicaclassica} applies. But in the quantum machine the set of physical properties that can be actually measured is drastically restricted, for only measurements of the form $e_u$ are allowed, with $u$ varying on the unit sphere (in particular, one cannot know the results of hidden measurements of the form $e_u^\lambda$). By referring to our mathematical representation, one restricts to a set ${\mathcal E}^{P} \subset {\mathcal E}$ of physical properties such that each $E \in {\mathcal E}^{P}$ is represented by a singleton $\left\{ u \right\}$, where $u$ is a point (equivalently, a vector) of $surf$, and $surf$ is a subset of the phase space of $P$. If one considers the language $L(x)$ in this specific case, such restriction implies considering a subset $\phi_{V}^{P}(x) \subset \phi_{V}(x)$ of practically testable wffs, hence a subset ${\mathcal P}_{V}^{P} \subset {\mathcal P}_{V}$ of practically testable physical propositions. The rules for assigning a probability to every pair consisting of a state and a physical property then imply that each physical proposition of ${\mathcal P}_{V}^{P}$ is represented by a vector on the unit sphere, that the ordered structure $({\mathcal P}_{V}^{P}, \subset)$ is an (elementary) modular lattice, and that the mapping which associates every vector on the sphere with the vector pointing in the opposite direction induces a complementation ${}^{\perp}$ on $({\mathcal P}_{V}^{P}, \subset)$. It follows that the structure $({\mathcal P}_{V}^{P}, \subset, {}^{\perp})$ (equivalently, $(\phi_{V}^{P}(x)/_{\equiv}, \le', {}^{\perp'})$)
But Aerts introduces hidden variables (the parameter $\lambda$) in his measurement processes. Hence the observative sublanguage $L$ of the general language $\mathcal L$ describing this class of systems (Sec. \ref{intro}) must contain terms interpreted on \emph{unsharp properties} (or \emph{effects}), which can be considered in CM as derived entities, defined by generalizing the standard notion of physical property. Thus one can construct $L$ by enlarging $L(x)$ by means of predicates denoting unsharp properties (for which Axiom CM2 does not hold, hence the physical preorder does not coincide with the logical preorder in $L$). It is then apparent that Aerts considers as verifiable sentences of $L$ only sentences that are logically equivalent to atomic sentences containing predicates that denote unsharp properties. This implies that the concrete logic that is obtained in this case is not a Boolean algebra. Indeed, Aerts' results entail that it is isomorphic to the QL of a spin--$\frac{1}{2}$ quantum particle, hence to the modular orthocomplemented lattice of all subspaces of ${\mathbb C}^2$.

Our informal discussion of this example is thus completed. We add that Aerts and his coworkers have constructed similar models for arbitrary quantum entities \cite{a85,a86,a87} which can be used to illustrate further our procedures.

%(ii) \emph{Kirkpatrick's playing cards} \cite{k03,k04}.

\section{Quantum mechanics and quantum logic\label{meccanicaquantistica}}
We intend to show in this section that standard QL can be recovered (up to an equivalence relation) as the concrete logic of QM within the general classical approach sketched in Secs. \ref{linguaggio} and \ref{physics}. To this end we preliminarily state the following fundamental axiom of QM, which establishes the mathematical representation of states and physical properties in QM in the case of quantum systems in which superselection rules do not occur.
%\begin{figure}
%\begin{center}
%\includegraphics[scale=0.6]{fig4}
%\end{center}
%\emph{Fig. 4.}
%\end{figure}

\vspace{.2cm}
\noindent
\textbf{Axiom QM1.} \emph{Every physical system $\Omega$ is represented in QM by a triple $({\mathscr H}, \psi, \omega)$, where ${\mathscr H}$, $\psi$ and $\omega$ are defined as follows.} \\ 
\emph{$\mathscr H$ is a Hilbert space over the complex field $\mathbb C$ associated with the physical system $\Omega$.} \\
\emph{$\psi: S \in {\mathscr S} \longmapsto \psi(S) \in {\mathcal L}_{1}({\mathscr H})$,} \\
\emph{where ${\mathcal L}_{1}({\mathscr H})$ is the set of all one--dimensional subspaces of $\mathscr H$.} \\
\emph{$\omega: E \in {\mathcal E} \longmapsto \omega(E) \in {\mathscr L}({\mathscr H})$,} \\
\emph{where ${\mathscr L}({\mathscr H})$ is the set of all closed subspaces  of $\mathscr H$.} \\
\emph{The mappings $\psi$ and $\omega$ are bijective.}

\vspace{.2cm}
\noindent
The following statement is then well known in QM.

\vspace{.2cm}
\noindent
\textbf{Prop. 6.1.} \emph{The structure $({\mathscr L}({\mathscr H}), \subset, {}^{\underline{\perp}})$, where ${}^{\underline{\perp}}$ is the mapping which associates every closed subspace of ${\mathscr L}({\mathscr H})$ with its orthogonal complement, is a complete orthomodular lattice, which is canonically equivalent to the algebra $({\mathscr L}({\mathscr H}), \Cap, \Cup, {}^{\underline{\perp}})$ (where $\Cap$ and $\Cup$ denote the meet and join, respectively, in the lattice $({\mathscr L}({\mathscr H}), \subset, {}^{\underline{\perp}})$).}

\vspace{.2cm}
\noindent
The lattice $({\mathscr L}({\mathscr H}), \subset, {}^{\underline{\perp}})$ or, equivalently, the algebra $({\mathscr L}({\mathscr H}), \Cap, \Cup, {}^{\underline{\perp}})$, plays a crucial role in standard QL. Indeed it allows us to define a lattice structure on ${\mathcal E}$ by means of the following definition.

\vspace{.2cm}
\noindent
\textbf{Def. 6.1.} \emph{We call} lattice of physical properties \emph{in QM the orthomodular lattice $({\mathcal E}, \angle, {}^{\underline{\perp}})$, where $\angle$ denotes the order and ${}^{\underline{\perp}}$ the orthocomplementation induced on ${\mathcal E}$, via $\omega$, by the order $\subset$ and the orthocomplementation  ${}^{\underline{\perp}}$, respectively, defined on ${\mathscr L}({\mathscr H})$, and denote by $({\mathcal E}, \Cap, \Cup, {}^{\underline{\perp}})$ the algebra canonically equivalent to $({\mathcal E}, \angle, {}^{\underline{\perp}})$ and isomorphic to $({\mathscr L}({\mathscr H}), \Cap, \Cup, {}^{\underline{\perp}})$.}

\vspace{.2cm}
\noindent
We can now introduce a further lattice in QM by means of the following definition, which is standard in the foundations of QM \cite{bc81}.

\vspace{.2cm}
\noindent
\textbf{Def. 6.2.} \emph{Let $E \in \mathcal E$. We call} certainly yes domain \emph{of $E$ the subset of states}
\begin{center}
\emph{${\mathscr S}_{E}=\{ S \in {\mathscr S} \ | \ \psi(S) \subset \omega(E) \}$ }
\end{center}
\emph{and put ${\mathcal P}_{{\mathcal E}}=\{ {\mathscr S}_{E} \ | \ E \in {\mathcal E} \}$.}

\vspace{.2cm}
\noindent
Indeed, the following proposition holds.

\vspace{.2cm}
\noindent
\textbf{Prop. 6.2.} \emph{The mapping}
\begin{center}
\emph{$\rho: {\mathscr S}_{E} \in {\mathcal P}_{{\mathcal E}} \longmapsto  \omega(E) \in {\mathscr L}({\mathscr H})$}
\end{center}
\emph{is bijective and, for every $E, F \in {\mathcal E}$, ${\mathscr S}_{E} \subset {\mathscr S}_{F}$ iff $\omega(E) \subset \omega(F)$. Therefore the structure $({\mathcal P}_{{\mathcal E}}, \subset, {}^{\underline{\perp}})$, where ${}^{\underline{\perp}}$ denotes the orthocomplementation induced on ${\mathcal P}_{{\mathcal E}}$, via $\rho$, by the orthocomplementation ${}^{\underline{\perp}}$ defined on ${\mathscr L}({\mathscr H})$, is an orthomodular lattice isomorphic to $({\mathscr L}({\mathscr H}), \subset, {}^{\underline{\perp}})$. Then the algebra $({\mathcal P}_{{\mathcal E}}, \Cap, \Cup, {}^{\underline{\perp}})$ canonically equivalent to $({\mathcal P}_{{\mathcal E}}, \subset, {}^{\underline{\perp}})$ is isomorphic to $({\mathscr L}({\mathscr H}), \Cap, \Cup, {}^{\underline{\perp}})$.}

\vspace{.2cm}
\noindent
We have thus singled out three isomorphic lattices, that is, $({\mathscr L}({\mathscr H}), \Cap, \Cup, {}^{\underline{\perp}})$, $({\mathcal E}, \Cap, \Cup, {}^{\underline{\perp}})$ and $({\mathcal P}_{{\mathcal E}}, \Cap, \Cup, {}^{\underline{\perp}})$, which only differ because of their supports and can be identified with the standard QL in the literature (which takes its name from the interpretation of the elements of ${\mathscr L}({\mathscr H})$ proposed by Birkhoff and von Neumann).% In particular, we can recover $({\mathcal P}_{{\mathcal E}}, \Cap, \Cup, {}^{\underline{\perp}})$ as the algebraic structure associated with the concrete logic of QM in our approach proceeding as follows. 

%First of all, we state the following axiom which relates the mathematical representation in Axiom QM1 with the set--theoretical representation of states and physical properties provided in Sec. \ref{linguaggio}.

We introduce now two axioms that can be justified in QM on the basis of the intended interpretation provided in Sec. \ref{linguaggio}. The first of them relates the mathematical representation in Axiom QM1 with the set--theoretical representation of states and physical properties provided in Sec. \ref{linguaggio}.

\vspace{.2cm}
\noindent
\textbf{Axiom QM2.} \emph{For every $E \in {\mathcal E}$, $S \in {\mathscr S}$,}
\begin{center}
\emph{$\psi(S) \in \omega(E)$ (equivalently, $S \in {\mathscr S}_{E}$) iff $ext(S) \subset ext(E)$,} \\
\emph{$\psi(S) \in \omega(E^{\underline{\perp}})$ (equivalently, $S \in {\mathscr S}_{E^{\underline{\perp}}}$) iff $ext(S) \subset {\mathcal U} \setminus ext(E)$.}
\end{center}

\vspace{.2cm}
\noindent
\emph{Physical justification}. For every physical property $E$ there exist in QM infinitely many states such that the quantum probability of $E$ in each of them is neither 0 nor 1. Hence there exist preparing devices that can be used to prepare ensembles of physical objects such that, for every ensemble, some elements display $E$ whenever a measurement of $E$ is performed and some do not. It follows that one cannot assert in QM that, for every state $S$ and property $E$, a physical object prepared by a preparing device belonging to $S$ either possesses or does not possess the property $E$. Therefore one can only characterize the sets of states for which one of the two alternatives occur.

The last axiom then distinguishes the notion of verifiability in QM from the notion of verifiability in CM.

\vspace{.2cm}
\noindent
\textbf{Axiom QM3.} \emph{The set $\phi_{V}(x)$ of all verifiable wffs of $\phi(x)$ is strictly included in $\phi(x)$.}

\vspace{.2cm}
\noindent
\emph{Physical justification}. A nontrivial compatibility relation exists in QM which prohibits testing sentences as $E(x) \land F(x)$ if the physical properties $E$ and $F$ are not compatible.

By using Axiom QM2 one can easily prove the following statement.

\vspace{.2cm}
\noindent
\textbf{Prop. 6.3.} \emph{For every $\alpha(x) \in \phi_{V}(x)$ and $S \in {\mathscr S}$,}
\begin{center}
\emph{$\alpha(x)$ is C--true (C--false) in $S$ iff $S \in {\mathscr S}_{E_{\alpha}}$ ($S \in ({\mathscr S}_{E_{\alpha}})^{\underline{\perp}}$).}
\end{center}

\vspace{.2cm}
\noindent
We can now introduce quantum connectives as operations on $\phi_{V}(x)$, as follows.

\vspace{.2cm}
\noindent
\textbf{Def. 6.3.} \emph{We call} quantum negation\emph{,} quantum meet \emph{and} quantum join \emph{the operations ${}^{\perp_{Q}}$, $\land_{Q}$ and $\lor_{Q}$, respectively, on $\phi_{V}(x)$, defined as follows.}

\emph{${}^{\perp_{Q}}: \alpha(x) \in \phi_{V}(x) \longmapsto {\alpha}^{\perp_{Q}}(x)={E}_{\alpha}^{\underline{\perp}}(x) \in \phi_{V}(x)$,}

\emph{$\land_{Q}: (\alpha(x), \beta(x)) \in \phi_{V}(x) \times \phi_{V}(x) \longmapsto  \alpha(x)  \land_{Q}  \beta(x)=(E_{\alpha} \Cap E_{\beta})(x) \in \phi_{V}(x)$,}

\emph{$\lor_{Q}: (\alpha(x), \beta(x)) \in \phi_{V}(x) \times \phi_{V}(x) \longmapsto  \alpha(x)  \lor_{Q}  \beta(x)=(E_{\alpha} \Cup E_{\beta})(x) \in \phi_{V}(x)$.}

\vspace{.2cm}
\noindent
The proof of the following statements is then straightforward.

\vspace{.2cm}
\noindent
\textbf{Prop. 6.4.} \emph{(i) For every $\alpha(x) \in \phi_{V}(x)$, $p_{\alpha}= {\mathscr S}_{E_\alpha}$, hence ${\mathcal P}_{V}={\mathcal P}_{{\mathcal E}}$.}

\emph{(ii) For every $E, F \in {\mathcal E}$,}
\begin{center}
\emph{$E=F$ iff $E(x) \equiv F(x)$ iff $E(x) \approx F(x)$,}
\end{center}
\emph{hence the equivalence relations $\equiv$ and $\approx$ coincide on $\phi_{V}(x)$.} 

\emph{(iii) The quantum negation ${}^{\perp_{Q}}$ is a weak complementation on $(\phi_{V}(x), \prec)$, hence $(\phi_{V}(x), \prec, {}^{\perp_{Q}})$ is the concrete logic of QM.}

\emph{(iv) For every $\alpha(x) \in \phi_{V}(x)$, the equivalence class $[\alpha(x)]_{\approx}$ contains one and only one wff of ${\mathcal E}(x)$, that is, $E_{\alpha}(x)$, hence the mapping}

\emph{${}^{\perp_{Q}'}: [\alpha(x)]_{\approx} \in \phi_{V}(x)/_{\approx} \longmapsto [\alpha(x)]_{\approx}^{\perp_{Q}'}=[\alpha^{\perp_{Q}}(x)]_{\approx}=[E_{\alpha}^{\underline{\perp}}(x)]_{\approx} \in \phi_{V}(x)/_{\approx}$}

\emph{is well defined and bijective.}

\emph{(v) The mapping}

\emph{$\xi: [\alpha(x)]_{\approx} \in \phi_{V}(x)/_{\approx} \longmapsto p_{\alpha} \in {\mathcal P}_{{\mathcal E}}$}

\emph{is an order isomorphism of $(\phi_{V}(x)/_{\approx}, \prec', {}^{\perp_{Q}'})$ onto $({\mathcal P}_{{\mathcal E}}, \subset, {}^{\underline{\perp}})$ which maps $[\alpha(x)]_{\approx}^{\perp_{Q}'}$ on $p_{\alpha}^{\underline{\perp}}$, hence $(\phi_{V}(x)/_{\approx}, \prec', {}^{\perp_{Q}'})$ is a standard QL.} 

\emph{(vi) Let $\land_{Q}'$ and $\lor_{Q}'$ be the operations induced by $\prec'$ on $\phi_{V}(x)/_{\approx}$. The, for every $\alpha(x), \beta(x)\in \phi_{V}(x)$,}

\emph{$[\alpha(x) \land_{Q} \beta(x)]_{\approx}=[(E_{\alpha} \Cap E_{\beta})(x)]_{\approx}=[\alpha(x)]_{\approx}\land_{Q}' [\beta(x)]_{\approx}=\xi^{-1}(p_{\alpha} \Cap p_{\beta})$,}

\emph{$[\alpha(x) \lor_{Q} \beta(x)]_{\approx}=[(E_{\alpha} \Cup E_{\beta})(x)]_{\approx}=[\alpha(x)]_{\approx}\lor_{Q}' [\beta(x)]_{\approx}=\xi^{-1}(p_{\alpha} \Cup p_{\beta})$.}

\vspace{.2cm}
\noindent
Prop. 6.4, (v), shows that a standard QL can be recovered as the orthomodular lattice associated with the concrete logic of QM in our classical framework. Prop. 6.4, (vi), shows that the concrete logic $(\phi_{V}(x), \prec, {}^{\perp_{Q}})$ is canonically equivalent to the algebraic structure $(\phi_{V}(x), {}^{\perp_{Q}}, \land_{Q}, \lor_{Q})$. But it must be stressed that the wffs of $\phi_{V}(x)$ play the role of atomic wffs in the concrete logic $(\phi_{V}(x), {}^{\perp_{Q}}, \land_{Q}, \lor_{Q})$, and that the meaning of the derived and theory--dependent quantum connectives (which is determined by QM) must be clearly distinguished from the meaning of classical connectives in our approach, in agreement with the known principle of Quine, ``change of logic, change of subject'' \cite{g92,q06}.

We have thus accomplished our task. We recall however from Sec. \ref{intro} that our procedure for recovering standard QL can be charged to be purely formal if one adopts the orthodox interpretation of QM. To overcome this objection we have observed that a physical interpretation of our procedure can be given by referring to the recent proposal of a theory (\emph{ESR model}) which generalizes QM embedding its mathematical apparatus into a broader mathematical formalism and reinterpreting quantum probabilities as conditional on detection rather than absolute \cite{gs09b,gs08,gs10a}. We can be more precise here and specify that objectivity of physical properties, which holds in the ESR model, allows one to provide, for every $E \in {\mathcal E}$, a physical interpretation of $ext(E)$ as the set of all physical objects that possess the property $E$ independently of any measurement. This interpretation is obviously impossible if the orthodox interpretation of quantum probabilities is maintained.

\section{The orthodox approach to standard quantum logic\label{quantum_logic}}
Because of the objection discussed at the end of Sec. \ref{meccanicaquantistica}, orthodox quantum logicians avoid associating an extension with the predicates denoting physical properties.\footnote{Note that a similar objection does not occur in the case of predicates denoting states, because the extension $ext(S)$ of a state $S \in {\mathscr S}$ can be interpreted as the set of all physical objects that are actually prepared in the state $S$.} Hence they introduce standard QL by adopting procedures that are mainly based on the mathematical representation of states and effects in QM \cite{r98,dcgg04}. To allow a comparison with our approach we sketch in this section an introduction to standard QL which particularizes and simplifies the methods described in \cite{dcgg04}.

Let us consider a quantum system $\Omega$ and the notions of pure state, physical property and physical object as defined in QM \cite{bc81,l83}. Then, let us denote by $L_{Q}$ a formal language, intended to express basic notions and relations in QM, constructed as follows.

\vspace{.2cm}
\noindent
\textbf{Def. 7.1.} \emph{The} alphabet \emph{of $L_Q$ consists of the following elements.} \\
\emph{A set ${\mathcal E}= \{ E, F, \ldots \}$ of atomic sentences (intended interpretation: physical properties).} \\
\emph{Connectives $\lnot_{Q}, \land_{Q}, \lor_{Q}$.} \\
\emph{Parentheses $(\, , \,)$.}

\vspace{.2cm}
\noindent
\textbf{Def. 7.2.} \emph{The set $\phi_{Q}$ of all} well formed formulas \emph{(}wffs\emph{) of $L_{Q}$ is the set obtained by applying recursively standard} formation rules \emph{(to be precise, for every $E \in {\mathcal E}$, $E \in \phi_{Q}$; for every $\alpha \in \phi_{Q}$, $\lnot_{Q}\alpha \in \phi_{Q}$; for every $\alpha, \beta \in \phi_{Q}$, $\alpha \land_{Q} \beta \in \phi_{Q}$ and $\alpha \lor_{Q} \beta \in \phi_{Q}$).}

\vspace{.2cm}
\noindent
\textbf{Def. 7.3.} \emph{By referring to the algebraic structure $({\mathcal P}_{{\mathcal E}}, \Cap, \Cup, {}^{\underline{\perp}})$ introduced in Sec. \ref{meccanicaquantistica}, Prop. 6.2, a} physical proposition \emph{$p_{\alpha}$ is associated with every $\alpha \in \phi_{Q}$, recursively defined as follows.}

\emph{For every $E \in \mathcal E$, $p_E={\mathscr S}_E$.} 

\emph{For every $\alpha, \beta \in \phi_{Q}$, $p_{\lnot_{Q} \alpha}=(p_{\alpha})^{\underline{\perp}}$, $p_{\alpha \land_{Q} \beta}=p_{\alpha} \Cap p_{\beta}$, $p_{\alpha \lor_{Q} \beta}=p_{\alpha} \Cup p_{\beta}$.} 

\vspace{.2cm}
\noindent
The proof of the following statements is then straightforward.

\vspace{.2cm}
\noindent
\textbf{Prop. 7.1.} \emph{For every $\alpha \in \phi_{Q}$ there exists a unique $E_{\alpha} \in {\mathcal E}$ such that $p_{\alpha} =p_{E_{\alpha}}$, and the set ${\mathcal P}_{Q}$ of all physical propositions associated with wffs of $\phi_{Q}$ coincides with ${\mathcal P}_{{\mathcal E}}$.}

\vspace{.2cm}
\noindent
The notion of Q--truth can now be introduced by means of the following definition.

\vspace{.2cm}
\noindent
\textbf{Def. 7.4.} \emph{For every $\alpha \in \phi_{Q}$ and $S \in \mathscr S$, $\alpha$ is} Q--true \emph{(}quantum true\emph{) in $S$ iff $S \in p_{\alpha}$, $\alpha$ is} Q--false \emph{(}quantum false\emph{) in $S$ iff $S \in (p_{\alpha})^{\underline{\perp}}$, $\alpha$ has no Q--truth value (equivalently, $\alpha$ is} Q--indeterminate\emph{) iff $S \notin p_{\alpha} \cup (p_{\alpha})^{\underline{\perp}}$.}

\vspace{.2cm}
\noindent
One can then prove the following nontrivial statement.

\vspace{.2cm}
\noindent
\textbf{Prop. 7.2.} \emph{For every $\alpha \in \phi_{Q}$ and $S \in \mathscr S$,}
\begin{center}
\emph{$\alpha$ has a Q--truth value in $S$ iff $\psi(S) \subset \omega(E_{\alpha}) \cup (\omega(E_{\alpha}))^{\underline{\perp}}$ iff \\ the probability of the property $E_{\alpha}$ in the state $S$ is either 1 or 0 iff \\ a measurement of the property $E_{\alpha}$ exists which does not modify the state $S$.}
\end{center}

\noindent
\textbf{Proof.} The first equivalence follows from Def. 7.4, Def. 6.1 and Prop. 6.2. The second equivalence follows from the standard rules of QM. The third equivalence is proven in \cite{gs04}.

\vspace{.2cm}
\noindent
\textbf{Def. 7.5.} \emph{The binary relations of} quantum preorder \emph{$\le_{Q}$ and} quantum equivalence \emph{$\equiv_{Q}$ on $\phi_{Q}$ are defined by setting, for every $\alpha,\beta \in \phi_Q$,} 
\begin{center}
\emph{$\alpha \le_{Q} \beta$ iff (for every $S \in {\mathscr S}$, $\alpha$ is Q--true in $S$ implies $\beta$ is Q--true in $S$)} 
\end{center}
\emph{and}
\begin{center}
\emph{$\alpha \equiv_{Q} \beta$ iff (for every $S \in {\mathscr S}$, $\alpha$ is Q--true in $S$ iff $\beta$ is Q--true in $S$).}
\end{center}

\vspace{.2cm}
\noindent
The proof of the following statements is then straightforward.

\vspace{.2cm}
\noindent
\textbf{Prop. 7.3.} \emph{(i) For every $\alpha,\beta \in \phi_Q$,}
\begin{center}
\emph{$\alpha \le_{Q} \beta$ iff $p_\alpha \subset p_\beta$ iff (for every $S \in {\mathscr S}$, $\beta$ is Q--false in $S$ implies $\alpha$ is Q--false in $S$) iff $(p_\beta)^{\underline{\perp}} \subset (p_\alpha)^{\underline{\perp}}$}
\end{center}
\emph{and}
\begin{center}
\emph{$\alpha \equiv_{Q} \beta$ iff ($\alpha \le_{Q} \beta$ and $\beta \le_{Q} \alpha$) iff $p_\alpha=p_\beta$ iff (for every $S \in {\mathscr S}$, $\beta$ is Q--false in $S$ iff $\alpha$ is Q--false in $S$) iff $(p_\beta)^{\underline{\perp}}=(p_\alpha)^{\underline{\perp}}$.} 
\end{center}

\emph{(ii) The equivalence relation $\equiv_{Q}$ is compatible with $\lnot_{Q}$, $\land_{Q}$ and $\lor_{Q}$ (to be precise, for every $\alpha, \beta, \gamma, \delta \in \phi_Q$, $\alpha \equiv_{Q} \beta$ implies $\lnot_{Q}\alpha \equiv_{Q} \lnot_{Q}\beta$, $\alpha \equiv_{Q} \gamma$ and $\beta \equiv_{Q} \delta$ imply $\alpha \land_Q \beta \equiv_{Q} \gamma \land_Q \delta$ and $\alpha \lor_Q \beta \equiv_{Q} \gamma \lor_Q \delta$).} 

\emph{(iii) Let $\land_{Q}'$, $\lor_{Q}'$, $\lnot_{Q}'$ denote the operations canonically induced on $\phi_{Q}/_{\equiv_{Q}}$ by $\land_{Q}$, $\lor_{Q}$, $\lnot_{Q}$, respectively. Then, the mapping $\zeta: [\alpha]_{\equiv_Q} \in \phi_{Q}/_{\equiv_{Q}} \longmapsto p_{\alpha} \in {\mathscr S}_{E_\alpha}$ is an isomorphism of $(\phi_{Q}/_{\equiv_{Q}}, \land_{Q}', \lor_{Q}', \lnot_{Q}')$ onto $({\mathcal P}_{{\mathcal E}}, \Cap, \Cup, {}^{\underline{\perp}})$ (}standard QL\emph{).}

\vspace{.2cm}
\noindent
We have thus concluded our short presentation of the orthodox approach to QL. By comparing our general methods in Secs. \ref{linguaggio} and \ref{physics} with this approach we can single out some relevant similarities and differences.

(i) The language $L_{Q}$ is a propositional logic  in which no reference is done to individual examples of physical systems, and states are considered as possible worlds of a Kripkean semantics, not as predicates, in the orthodox approach.

(ii) The assignment of a Q--truth value to a wff $\alpha$ of $L_{Q}$ is given resorting to the physical proposition $p_{\alpha}$ associated with $\alpha$, hence it parallels the assignment of a C--truth value to a wff $\alpha(x)$ of $L(x)$ introduced in Def. 3.2 rather than the assignment of a truth value introduced in Def. 2.3.

(iii) Prop. 7.2 shows that the definition of Q--truth introduces a verificationist notion of truth, which is considered problematic by many logicians and philosophers \cite{r40,p61,p63,l00} and is at odds with our choice of introducing a notion of truth as correspondence in Secs. \ref{linguaggio} and \ref{physics}, carefully distinguishing between truth and verifiability. Consistently, the set ${\mathcal P}_{Q}$ of all physical propositions associated with wffs of $\phi_{Q}$ coincides with the set ${\mathcal P}_{{\mathcal E}}$ of all propositions associated with atomic wffs of $\phi_Q$, %(hence the structure $({\mathcal P}_{Q}, \Cap, \Cup)$ parallels the (order) structure $({\mathcal P}_{V}, \subset)$ introduced in Prop. 3.3.1 rather than $({\mathcal P}, \subset)$) 
which implies that only verifiable quantum sentences have been taken into account from the very beginning when constructing $L_{Q}$.

%The above similarities and differences are clarified and explained by our treatment in the next section, where standard QL is formally recovered within the classical framework worked out in Secs. \ref{linguaggio} and \ref{physics}. 

\section*{Acknowledgements}
The authors are greatly indebted with Carlo dalla Pozza and Marco Persano for reading the manuscript and providing useful remarks and suggestions.

%${}^{\underline{\perp}}: A \mapsto A^{\underline{\perp}}$


\begin{thebibliography}{99}
\bibitem{h34} Heyting, A. (1934). Matematische Grundlagenforschung, Intuitionismus, Beweistheorie. \emph{Ergebnisse der Matematik und ihrer Grenzgebiete}, \emph{3}, Berlin.

\bibitem{h56} Heyting, A. (1956). \emph{Intuitionism. An introduction}. Amsterdam: North-Holland.

\bibitem{l20} {\L}ukasiewicz, J. (1920). O logice tr\'{o}jwarto\'{s}ciowej. \emph{Ruch Filozoficzny}, \emph{5}, 169--171; English translation, On three-valued logic. Collected in Borkowski, L. (Ed.) (1970). \emph{Jan {\L}ukasiewicz, selected works}. 87--88. Amsterdam: North-Holland Publishing Company.

\bibitem{ab75} Anderson, A. R., \& Belnap, N. D. (1975). \emph{Entailment: the logic of relevance and necessity, Vol. I}. Princeton, NJ: Princeton University Press.

\bibitem{abd92} Anderson, A. R., Belnap, N. D., \& Dunn, J. M. (1992). \emph{Entailment: the logic of relevance and necessity, Vol. II}. Princeton, NJ: Princeton University Press.

\bibitem{g87} Girard, J. Y. (1987). Linear logic. \emph{Theoretical Computer Science} \emph{50}, 1--102.

\bibitem{t01} Tarski, A. (1933). Pojecie prawdy w jezykash nauk dedukcyjnych. \emph{Acta Towarzystwei Naukowego i Literakiego Warszawskiego}, \emph{34}, V--16; English translation, The concept of truth in formalized languages. Collected in Tarski, A. (1956). \emph{Logic, semantics, metamathematics}. 152--268. Oxford: Blackwell.

\bibitem{t02} Tarski, A. (1944). The semantic conception of truth and the foundations of semantics. \emph{Philosophy and phenomenological research}, \emph{IV}; Collected in Linsky, L. (Ed.) (1952). \emph{Semantics and the philosophy of language}. Urbana: University of Illinois Press.

\bibitem{h74} Haack, S. (1974). \emph{Deviant logic}. Cambridge: Cambridge University Press.

\bibitem{h78} Haack, S. (1978). \emph{Philosophy of logic}. Cambridge: Cambridge University Press.

\bibitem{dpg95} Dalla Pozza, C., \& Garola, C. (1995). A pragmatic interpretation of intuitionistic propositional logic. \emph{Erkenntnis} \emph{43}, 81--109.

\bibitem{j74} Jammer, M. (1974). \emph{The philosophy of quantum mechanics}. New York: Wiley-Interscience.

\bibitem{bvn36} Birkhoff, G., \& von Neumann, J. (1936). The logic of quantum mechanics. \emph{Annals of Mathematics}, \emph{37}, 823--843.

\bibitem{r98} R\'{e}dei, M. (1998). \emph{Quantum logic in algebraic approach}. Dordrecht: Kluwer.

\bibitem{dcgg04} Dalla Chiara, M. L., Giuntini, R., \& Greechie, R. (2004). \emph{Reasoning in quantum theory}. Dordrecht: Kluwer.

\bibitem{g92} Garola, C. (1992). Truth versus testability in quantum logic. \emph{Erkenntnis}, \emph{37}, 197--222.

\bibitem{dc00} Dalla Chiara, M. L. (1974). \emph{Logica}. Milano: ISEDI.

\bibitem{b66} Bell, J. S. (1966). On the problem of hidden variables in quantum mechanics. \emph{Review of Modern Physics}, \emph{38}, 447--452.

\bibitem{ks67} Kochen, S. \& Specker, E. P. (1967). The problem of hidden variables in quantum mechanics. \emph{Journal of Mathematics and Mechanics}, \emph{17}, 59--87.

\bibitem{b64} Bell, J. S. (1964). On the Einstein-Podolsky-Rosen paradox. \emph{Physics}, \emph{1}, 195--200.

\bibitem{gs96a} Garola, C., \& Solombrino, L. (1996). The theoretical apparatus of semantic realism: a new language for classical and quantum physics. \emph{Foundations of Physics}, \emph{26}, 1121--1164.

\bibitem{gs96b} Garola, C., \& Solombrino, L. (1996). Semantic realism versus EPR-like paradoxes: the Furry, Bohm-Aharonov, and Bell paradoxes. \emph{Foundations of Physics}, \emph{26}, 1329--1356.

\bibitem{gp04} Garola, C., \& Pykacz, J. (2004). Locality and measurements within the SR model for an objective interpretation of quantum mechanics. \emph{Foundations of Physics}, \emph{34}, 449--475.

\bibitem{gs09b} Garola, C., \& Sozzo, S. (2009). The ESR model: a proposal for a noncontextual and local Hilbert space extension of QM. \emph{Europhysics Letters}, \emph{86}, 20009.

\bibitem{gs08} Garola, C., \& Sozzo, S. (2010). Embedding quantum mechanics into a broader noncontextual theory: a conciliatory result. \emph{International Journal of Theoretical Physics}, \emph{49}, 3101--3117.

\bibitem{gs10a} Garola, C., \& Sozzo, S. (2011). Generalized observables, Bell's inequalities and mixtures in the ESR model. \emph{Foundations of Physics}, \emph{41}, 424--449.

\bibitem{bc81} Beltrametti, E. G., \& Cassinelli, G. (1981). \emph{The logic of quantum mechanics}. Reading, MA: Addison--Wesley.

\bibitem{l83} Ludwig, G. (1983). \emph{Foundations of quantum mechanics I}. Berlin: Springer.

\bibitem{a88} Aerts, D. (1988). The physical origin of the EPR paradox and how to violate Bell inequalities by macroscopic systems. In Lahti, P., \emph{et al.} (Eds.), \emph{Symposium on the foundations of modern physics}. 305-320. Singapore: World Scientific.

%Aerts, D., 1988b, The description of separated systems and a possible explanation for the probabilities of quantum mechanics, in Microphysical Reality and Quantum Formalism, eds. van der Merwe et al., Kluwer Academic Publishers.

\bibitem{a91} Aerts, D. (1991). A macroscopic classical laboratory situation with only macroscopic classical entities giving rise to a quantum mechanical probability model. In Accardi, L. (Ed.), \emph{Quantum probability and related topics}. 75--85. Singapore: World Scientific.

\bibitem{a95} Aerts, D. (1995). Quantum structures: an attempt to explain their appearance in nature. \emph{International Journal of Theoretical Physics}, \emph{34}, 1165--1186.

\bibitem{a98} Aerts, D. (1998). The hidden measurement formalism: what can be explained and where quantum paradoxes remain. \emph{International Journal of Theoretical Physics}, \emph{37}, 291--304.

\bibitem{a85} Aerts, D. (1985). A possible explanation for the probabilities of quantum mechanics and a macroscopic situation that violates Bell inequalities. In Mittelstaedt, P., \emph{et al.} (Eds.), \emph{Recent developments in quantum logic}. 235-251. Mannheim: Bibliographisches Institut. 

\bibitem{a86} Aerts, D. (1986). A possible explanation for the probabilities of quantum mechanics. \emph{Journal of Mathematical Physics}, \emph{27}, 202--210.

\bibitem{a87} Aerts, D. (1987). The origin of the non-classical character of the quantum probability model. In Blanquiere, A., \emph{et al.} (Eds.), \emph{Information, complexity and control in quantum physics}. 77-100. Wien--New York: Springer--Verlag.

%\bibitem{a99a} Aerts, D. (1999). Foundations of quantum physics: a general realistic and operational approach. \emph{International Journal of Theoretical Physics}, \emph{38}, 289--358.

%\bibitem{a99b} Aerts, D. (1999). Quantum mechanics: structures, axioms and paradoxes. In Aerts, D., \& Pykacz, J. (Eds.), \emph{Quantum physics and the nature of reality}. 141--205. Dordrecht: Kluwer. 

%\bibitem{k03} Kirkpatrick, K. A. (2003). Quantal behavior in classical probability. \emph{Foundations of Physics Letters}, \emph{16}, 199--224.

%\bibitem{k04} Kirkpatrick, K. A. (2004). Compatibility and probability. \emph{arXiv: quant-ph/0403007}.

\bibitem{q06} Quine, W. V. O. (2006). \emph{Philosophy of logic}. Cambridge, MA: Harvard University Press.

\bibitem{gs04} Garola, C., \& Sozzo, S. (2004). A semantic approach to the completeness problem in quantum mechanics. \emph{Foundations of Physics}, \emph{34}, 1249--1266.

%\bibitem{ga08} C. Garola, ``Physical propositions and quantum languages,'' \emph{Int. J. Theor. Phys.} \textbf{47}, 90--103 (2008).

\bibitem{r40} Russell, B. (1940). \emph{An inquiry into meaning and truth}. New York: W. W. Norton \& Company.

\bibitem{p61} Pap, A. (1961). \emph{An introduction to the philosophy of science}. New York: The Free Press.

\bibitem{p63} Popper, K. (1963). \emph{Conjectures and refutations}. London: Routledge and Kegan Paul.

\bibitem{l00} Lycan, W. (2000). \emph{Philosophy of language: a contemporary introduction}. London: Routledge.

\end{thebibliography}
\end{document}